\documentclass[a4paper,11pt]{article}
\usepackage{xy}

\begin{document}
\begin{flushright}
IFUP-TH 08/15\\
SISSA 54/2008/EP 
\end{flushright}
\vspace{0.2cm}
\begin{center}{ \LARGE \textbf{Homotopy, monopoles and 't Hooft tensor} \vspace{0.1cm}\\\textbf{ for generic gauge groups}}\end{center}
\vspace{0.3cm}
\begin{center}\begin{large}
A. Di Giacomo$ \footnote{adriano.digiacomo@df.unipi.it} $ ,\,
L. Lepori$  \footnote{lepori@sissa.it}$ ,\, F. Pucci$
\footnote{pucci@fi.infn.it}$\end{large}\end{center}
\begin{center}
\vspace{0.8cm} $^{1}$ \emph{Dip. di Fisica Universit$\grave{a}$ di Pisa and INFN-Sezione di Pisa}\\
Largo Bruno Pontecorvo 3, 56127 Pisa, Italy

\hspace{0.5cm}

 $^{2} $ \emph{International School for Advanced Studies (SISSA)\\and INFN-Sezione di Trieste}\\
Via Beirut 2-4, 34014 Trieste, Italy

\hspace{0.5cm}

$^{3}$ \emph{Dip. di Fisica Universit$\grave{a}$ di Firenze and INFN-Sezione di Firenze}\\
Via G.Sansone 1, 50019 Sesto Fiorentino, Italy

\end{center}
\hspace{0.2cm}
\begin{abstract}
We study monopoles and corresponding 't Hooft tensor in a generic gauge theory. This issue is relevant to the understanding of the color confinement in terms of dual symmetry.
\end{abstract}

\section{Introduction} \label{section1}

Any explanation of color confinement in terms of a dual symmetry,
requires the existence of field configurations with non trivial
spatial homotopy \begin{math}\Pi_2\end{math}. This amounts to extend
the formulation of the theory to a spacetime with an arbitrary but
finite number of line-like singularities (monopoles) \cite{Gukov:2006jk}.
\newline A prototype example of such configuration is the 't Hooft -
Polyakov monopole \cite{'tHooft:1974qc}\cite{d1} in the $SO(3)$
gauge theory interacting with a Higgs scalar in the adjoint color
representation. It is a static soliton solution made stable by its
non trivial homotopy.\newline In the ''hedgehog'' gauge  the i-th
color component of the Higgs field \begin{math}\phi ( r ) =
\phi^i( r ) \sigma_i
\end{math} at large distances has the form\begin{equation}\phi^i \simeq \frac{r^i}{|\, \vec{r}\, |}\end{equation}
and is a mapping of the sphere \begin{math}S_2\end{math} at
spatial infinity on $SO(3)/U(1)$, with non trivial homotopy. In
the unitary gauge where \begin{math}\frac{\phi^i}{| \phi\, |}=
\delta_3^i\, \sigma_3\end{math} is diagonal,
 a line singularity appears starting from the location of the monopole.\newline
The Abelian field strength of the residual $U(1)$ symmetry in the unitary gauge is given by
\begin{equation}F_{\mu \nu }= \partial_{\mu} A_{\nu}^{3} - \partial_{\nu} A_{\mu}^{3}\label{Abelian}\end{equation}
The monopole configuration has zero electric field
(\begin{math}F_{0 i}= 0 \end{math}) and the magnetic field
\begin{math}H_i = \frac{1}{2}\, \epsilon_{ijk} F_{jk}\end{math} is
the field of a Dirac monopole of charge 2
\begin{equation} \vec{H} = \frac{1}{g} \frac{\vec{r}}{4 \pi r^3}  \texttt{ + Dirac  String}\end{equation}
In a compact formulation, like is lattice, the Dirac string is invisible and a violation of Bianchi identity occurs
\begin{equation}\vec{\nabla}\cdot \vec{H} = \frac{1}{g}\, \delta^3 ( x )\end{equation}
More formally, one can define a covariant field strength
\begin{math}F_{\mu \nu}\end{math} which coincides, in the unitary gauge, with the abelian
field strength of the residual symmetry \cite{'tHooft:1974qc}
\begin{equation}
F_{\mu\nu}=Tr(\hat{\phi}\,
G_{\mu\nu})- \frac{i}{g}\, Tr \left(\hat{\phi}\, [D_{\mu}\hat{\phi},D_{\nu}\hat{\phi}]\right)\label{eq.4}\end{equation}
\vspace{0,2 cm}
 Here\begin{displaymath}\hat{\phi} = \sum \hat{\phi}^a T^a\, \, \hspace{0.5cm} G_{\mu \nu} = \sum G_{\mu \nu}^a T^a\end{displaymath}
\begin{displaymath}\hat{\phi}_a = \frac{\phi_a}{|\, \phi_a  |} \, \, \hspace{0.5cm} D_{\mu}\hat{\phi} = \partial_{\mu} \hat{\phi} + i g [ A_{\mu} , \hat{\phi} ]\end{displaymath}\\ \begin{math}T^a\end{math} are the group generators with normalization \begin{math}Tr( T^a T^b ) = \frac{1}{2}\, \delta^{a b}\end{math}. $ F_{\mu \nu}$ is known as 't Hooft tensor.
A magnetic current can be defined as
\begin{equation}j_{\mu} = \partial_{\nu}\widetilde{F}_{\mu \nu}\label{S}\end{equation}
A non zero value of it signals the violation of Bianchi
identities. Furthermore, the current as defined in eq.(\ref{S}) is
identically conserved
\begin{equation}\partial^{\mu} j_{\mu} = 0\label{S1}\end{equation}
The main feature of eq.(\ref{eq.4}) is that linear and bilinear
terms in
\begin{math}A_{\mu},A_{\nu}\end{math} cancel and one has
identically
\begin{equation}F_{\mu \nu} =  Tr( \partial_{\mu}( \hat{\phi} A_{\nu} ) - \partial_{\mu}( \hat{\phi} A_{\nu} ) - \frac{i}{g}\, \hat{\phi} [ \partial_{\mu} \hat{\phi} , \partial_{\nu} \hat{\phi} ] )\label{5}\end{equation}
In the unitary gauge, where $\hat{\phi}=(0,0,1)$ and
\begin{math}\partial_{\mu}\hat{\phi}=0\end{math}, it reduces to eq.(\ref{Abelian}). \\
In a theory with no Higgs field a 't Hooft tensor can be defined
by choosing\begin{equation}\phi = U(x) \sigma_3
U(x)^{\dagger}\label{effective}\end{equation} with $U(x)$ any
element of the group, for example the parallel transport to $x$
from a fixed arbitrary point at infinity.  $U(x)^{\dagger}$ is the
gauge transformation to the unitary gauge.\newline Again a
conserved magnetic current, identifying a dual symmetry, can be
defined. In principle any field
$\phi$ in the adjoint representation can be used as effective
Higgs: all of them have the form of eq.(\ref{effective}) except
for a finite number of singularities and differ from each other by
a gauge transformation defined everywhere except at
singularities.\newline The generalization to $SU(N)$ is designed
in ref.\cite{'tHooft:1974qc} and developed in detail in
ref.\cite{a}. The strategy is to ask what fields $\phi$ would
allow the definition of 't Hooft tensor, with the cancelations
bringing from eq.(\ref{eq.4}) to eq.(\ref{5}), so that it becomes
the abelian residual field strength  in the unitary gauge.\newline The
answer is that there are $N-1$ such fields (as many as the rank of
the group), one for each fundamental weight. Explicitly
\begin{equation}\phi^a(x) = U(x) \phi^a_{0}U^{\dagger}(x)\end{equation}
with $\phi_0^a$ the fundamental weight
\begin{equation}
\phi_{0}^{a}=\frac{1}{N}\, \texttt{diag}\, (\, \underbrace{N-a,\ldots,N-a}_{a}\, , \underbrace{ - a ,\,\ldots , - a
}_{N-a})\label{eq:7}\end{equation} $(a = 1\ldots N-1)$. The invariance group of
$\phi^a_{0}$ is
\begin{math} SU(a)\times SU(N-a)\times U(1) \end{math}
\vspace{0.1cm} and the quotient group
\begin{math}\frac{SU(N)}{SU(a)\times SU(N-a) \times U(1)}\end{math} has non
trivial homotopy
\begin{displaymath}\Pi_2 \left( \frac{SU(N)}{SU(a)\times SU(N-a) \times U(1)} \right) =  \textbf{Z}\end{displaymath}
For a more precisely formulation see section 3 below.
There exist $N-1$ monopole species for $SU(N)$, one for each
$a$.\newline To connect with the approach of ref.\cite{ada2}, if
$\psi(x)$ is a generic hermitian operator in the adjoint
representation, it can be diagonalized to $\psi_{0}(x)$. Since the
maximal weights are a complete set of traceless $N\times N$
diagonal matrices, one has
\begin{equation}\psi_0(x)=  \sum_{a=1}^{N-1} c_{a}(x)
\phi^{a}_{0}\end{equation} and
\begin{equation}\psi(x)=\sum_{a=1}^{N-1} c_{a}(x)
\phi^{a}(x)\end{equation} $c_{a}(x)$ is the difference of two
subsequent eigenvalues of $\psi$
\begin{displaymath}c_a(x)= \psi_{0}(x)_a^{a} - \psi_{0}(x)_{a+1}^{a+1}\end{displaymath}
and is equal to zero at the sites where two eigenvalues coincide
and there a singularity appears in the unitary gauge,
corresponding to a monopole of species $a$ sitting at $x$.
\newline Recently some special groups like $G_2$ and $F_4$ became
of interest, since they have no center and seem to confine
\cite{berna1}, in contrast with the idea that center vortices
could be the configurations responsible for confinement
\cite{ad1}. It is thus interesting to investigate monopole
condensation in these systems.\newline However, for the group
$G_2$ and $F_4$ it proves impossible to construct a 't Hooft
tensor of the form of eq.(\ref{eq.4}): no solution exists for
$\phi^a$, such that eq.(\ref{eq.4}),(\ref{5}) are valid. Still, as
we shall see in the following, there are monopoles in these
theories and it is possible to define magnetic conserved currents.
The approach sketched above which works for $SU(2)$ and $SU(N)$
has to be modified for a more general construction of a 't Hooft
like tensor. We approach and solve this problem in the present
paper.

\section{Monopoles}

Let $G$ be a gauge group, which we shall assume to be compact and
simple. To define a monopole current we have to isolate an $SU(2)$
subgroup, and break it to its third component, say $T_3$.
This will be done by some "Higgs field" $\phi$  in the adjoint representation.\\
Our notation is the familiar one (see e.g.
\cite{Libro2}\cite{Libro1}). There are $r$ commuting generators of
$G$ ($r$=rank of group) which we shall denote as $H_i$
($i=1,..,r$). The other generators occur in pairs with opposite
values of Cartan eigenvalues:
\begin{equation}
\begin{array}{cc}
[H_{i},H_{j}\,]=0 & [H_{i},E_{\pm\, \vec{\alpha}}\, ]=\pm\, \alpha_i E_{\pm\, \vec{\alpha}}\\
\, & \, \\
{}[E_{\, \vec{\alpha}},E_{\vec{\beta}}\, ] = N_{\vec{\alpha},\,
\vec{\beta}}\, E_{\vec{\alpha} + \vec{\beta}} &
[E_{\vec{\alpha}},E_{- \vec{\alpha}}\,
]=\alpha_{i}H_{i}\end{array}\label{eq:5.2}\end{equation} where
$\vec{\alpha}=(\alpha_1 \ldots \alpha_r)$ and $N_{\alpha,\, \beta}
\neq 0 $ only if $\vec{\alpha} + \vec{\beta}$ is a root. The root
$\vec{\alpha}$ can be taken positive  ($-\, \vec{\alpha}$
negative). By definition, a root is positive if its first nonzero
component is positive: either $\vec{\alpha}$ or $- \vec{\alpha}$
is positive. Of course the choice is conventional and also depends
on the choice for the order of components. A positive root is
called simple if it cannot be written as the sum of two other
positive roots.\\
The way to associate an $su(2)$ algebra to each root is a trivial
renormalization of $E_{\pm \vec{\alpha}}$ . Defining
\begin{center}$\begin{array}{ccc}
T_{\pm}^{\alpha}= \sqrt{\frac{2}{(\vec{\alpha}\cdot
\vec{\alpha})}} E_{\pm \vec{\, \alpha}} & \,\, & T_{3}^{\alpha} =
\frac{\vec{\alpha}\cdot \vec{H}}{(\vec{\alpha}\cdot
\vec{\alpha})}\end{array}\label{eq:5.6}$\end{center} we have
\begin{center}$\begin{array}{ccc}
[T_{3}^{\alpha}, T_{\pm}^\alpha]=\pm\, T_{\pm}^{\alpha} & \,\, &
[T_{+}^{\alpha}, T_{-}^\alpha]= 2\,
T_{3}^{\alpha}\end{array}\label{eq:5.7}$\end{center} A Weyl
transformation is an invariance transformation of the algebra
which permutes the roots \cite{Libro2}\cite{Libro1}. It can be
proved that any root can be made a simple root by a Weyl
transformation (\cite{Libro2} III.10 pg.51). Furthermore it can
also be proved that the Weyl transformations are induced by
transformations of the group G  (\cite{Libro1} VIII.8  pg.193). If
the Higgs potential is invariant under G, we can then consider
without loss of generality only the $SU(2)$ subgroups related to
the simple roots.\\
A $vev$ of the field $\phi$ proportional to any of the fundamental
weights $\mu^i$, $i=(1,\ldots,r)$, corresponding to the i-th simple
root, identifies a monopole\footnote{This kind of breaking is
called \emph{maximal} and identifies $r$ magnetic charges, one for
each fundamental weight. Configurations carrying a non zero value
of more than one of this charge (non maximal breaking) exist
\cite{i}, but they don't add any new information concerning the
symmetry.}. Indeed recall that
\begin{center}$\begin{array}{ccc}
\mu^{i}= \vec{c}^{\, \, i} \cdot \vec{H}  & \,\, &  [\mu^{i},\,
T_{\pm}^{j}] = \pm\, \vec{c}_{i} \cdot \vec{\alpha}_{j}\,
T_{\pm}^{j} = \pm \, \delta_{ij} \, T_{\pm}^{j}
\end{array}\label{eq:5.8}$\end{center}
Taking
\begin{equation}\phi^i = T_{3}^{\, i} + ( \mu^{\, i} - T_{3}^{\, i} )\label{T3}\end{equation}
the last term commutes with $T_{\pm} ^{\, i}$, $T_{3} ^{\, i}$
\begin{equation} [\, \mu^{\, i},\, T_{\pm}^{\, j}\, ] = \pm\, \delta_{ij}\, T_{\pm}^{\, j}\, \, \, \, \, \, \, \, \, \, \, [\, \mu^{\, i},\, T_{3}^{\, j}\, ] = 0\, \, \, \, \, \, \, \, \, \, [T_{3}^{\, i},  \, T_{\pm}^{\, j}\, ] = \pm\, \delta_{ij}\, T_{\pm}^{\, j}\end{equation}
We can then write down explicitly the monopole solution by use of
the 't Hooft-Polyakov ansatz \cite{'tHooft:1974qc}\cite{d1} inside
the $SU(2)$ subgroup generated by $T_{\pm} ^{\, i}$, $T_{3}^{\,
i}$.
\begin{equation}A_{k}^i=A_k^a(\vec{r})T_a^{\, i},\qquad \phi(\vec{r})^i=\chi^a(\vec{r})T_a^{i}+
(\mu^{\, i} - T_3^{\, i}),\label{eq.12b}\end{equation}
where
\begin{equation}
A_{k}^{a}(\vec{r})=\epsilon_{akj}\frac{r^{j}}{r^{2}}, \qquad
\chi^a(\vec{r})=\frac{r^a}{r}\, \chi(r) \qquad \chi(\infty)=1
\label{eq.12c}
\end{equation}
The index $a$ indicates color, while the indices $k, j$ space
directions and the index $i$ refers to the root chosen. The little
group of $\phi$, $\tilde{H}$, is the product of the $U(1)$ generated
by $\mu^{\, i}$ times a group $H$ which has as Dynkin diagram the
diagram (connected or not connected) obtained by erasing from the
diagram of $G$ the root $\alpha_i$ and the links which connect it to
the rest (Levi subgroup):\begin{equation} \tilde{H}=H\times U(1)
\end{equation} Indeed $\phi=\mu^{\, i}$ commutes with
all the roots different from $\alpha_i$ and of course with the
$H_i$. The 't Hooft tensor will be, in the unitary gauge in which
$\phi^i$ is diagonal,
\begin{equation}F_{\mu \nu }^i= \partial_{\mu} A_{\nu}^{3} - \partial_{\nu} A_{\mu}^{3}\label{Abelian F.S}\end{equation}
where 3 labels the component along $T_{3}^{ i}$, the diagonal generator of
the broken $SU(2)$. It is straightforward to see from
eq.(\ref{Abelian F.S}) that the i-th monopole is charged under the
magnetic $U(1)$ generated by $T_{3}^{ i}$.

\section{Monopole charge and homotopy}
Monopole configurations can be classified in terms of the second
homotopy group $\Pi_2(G/\tilde{H})$.
In the following we will use the relationship \cite{c}
\begin{equation} \pi_{2}(G/\tilde{H})\simeq \texttt{ker}[\pi_1(\tilde{H}) \to\pi_1(G) ] \end{equation}
and we will compute $\pi_1(\tilde{H})$ following the formulation of \cite{Wu:1975es}. 
We consider two gauge fields, respectively
defined on north ($0\leq \theta<\pi/2$) and south ($\pi/2<\theta
\leq \pi$) hemisphere, of the form
\begin{equation}A_{\varphi}^{\pm}=\, \pm\, g\, T_3 (1 \mp \cos{\theta}),\end{equation}
with $\varphi$ the azimuthal direction.
 $A_\varphi^{+}$ is defined on the north hemisphere and
$A_\varphi^{-}$ in the south one.  $T_3$ is the third
component of the broken $SU(2)$.\\
On the equator this two solutions
must be transformed one into each other by a gauge transformation of the form
\begin{equation}\Omega = \texttt{exp}( i\, 2\, e\,  g\, T_3\, \varphi )\label{Omega}\end{equation}
which is single-valued if
\begin{equation} \texttt{exp}( i\, 4\, \pi\, e\,  g\,  T_3 ) = 1\label{Omega2}\end{equation}
In the simple case of $G = SU(2)$ and $\tilde{H} = U(1)$,
eq.(\ref{Omega2}) gives the Dirac quantization condition
\begin{equation}g = \frac{n}{2 e}\end{equation}
Monopoles are identified by an integer $n$, the winding number on
$\tilde{H} = U(1)$ group. Indeed
\begin{equation}\Pi_2(SU(2)/U(1)) = \Pi_1(U(1)) = \textbf{Z}\end{equation}
For a generic gauge group $G$ the discussion turns out to be more
involved, since the analysis of $\Pi_2(G/\tilde{H})$ is related to
the global (topological) structure of $G$ and $\tilde{H}$ which in
general cannot be inferred from their Lie algebras.\newline
In general
\begin{equation} \tilde{H}= \frac{H \times U(1)}{Z}\label{H}\end{equation}
where $Z$ is a subgroup of the center of $H \times U(1)$. This happens
when the identity of $G$ can be written not only as the identity of
$H$ times the identity of $U(1)$ but also as an element of $U(1)$
times a non trivial element $z$ of $H$. Since $U(1)$ commutes with
$H$, $z$ must commute with all elements of $H$ and hence it
belongs to the center of H. Mathematically speaking, $Z$ is the
kernel of the map $\Phi : \, H \times U(1) \to G$. \newline For
example, for $G = SU(N)$, one can check that the residual
invariance group is
\begin{equation}\tilde{H} = \frac{SU(a)\times SU(N-a)\times U(1)}{\textbf{Z}_{k}}\end{equation}
where $k$ is the $mcm$ between $a$ and $N-a$. The third component
of the broken SU(2) is
\begin{equation}T_3 = (0,\ldots ,1,-1,\ldots ,0),\end{equation}
so that the usual Dirac quantization $g = \frac{n}{2 e}$, in terms
of the minimal electric charge \cite{b}\cite{Preskill}, follows
from  eq.(\ref{Omega2}). Monopole configurations are labeled by
an integer $n$.\newline To see the correspondence between the $U(1)$
magnetic charges and the non-contractible loops on $\tilde{H}$, we
substitute the value of $e g$ as determined from eq.(\ref{Omega2})
into eq.(\ref{Omega}) obtaining \begin{equation}\Omega = \,
exp(i\, \, n\, T_3\, \varphi )\end{equation} Magnetic charges
(with various $n$) are associated to loops that wind $n$-times on
magnetic $U(1)$ , the subgroup generated by $T_3$.\newline From
the point of view of the $\tilde{H}$ group, every monopole charge
is in one-to-one correspondence with a loop that starts from
identity, moves inside the $U(1)$ to an element of the center of
$SU(a)\times SU(N-a)$  and comes back to identity along a path
into $SU(a)\times SU(N-a)$.\newline Algebraically one can write
(see eq.(\ref{T3})) \cite{b}\cite{Preskill}:
\begin{equation}T_3 = \phi + h \label{peso} \end{equation}
with
\begin{equation}\phi= \left(\, \frac{1}{a}\,,\, \, \cdots \,\, \frac{1}{a}\,,\,  - \frac{1}{N-a}\,,\, \, \cdots \,\, - \frac{1}{N-a}\,\right)\end{equation}
\begin{equation}h= \left(\, - \frac{1}{a}\,,\,\cdots \,  \frac{a-1}{ a}\,,\, \, \frac{a+1-N}{N-a}\,,\, \cdots \, \frac{1}{N-a}\,\right)\end{equation}
where $\phi$ is the effective Higgs and $h$ is an element of the
Cartan subalgebra of $H$. By use of formula (\ref{peso}), we
easily recognize that the loops in the $U(1)$ with winding number
$L$ correspond to magnetic charges $n = L\, k$ since, for $\varphi
= 2\pi$, $e^{i\, 2 \pi\, \phi\, L\, k} = I$. Charges of the form
\begin{equation}n = q +  L\, k \qquad q \neq0\end{equation}
are associated to loops that go inside $U(1)$ from identity to
\begin{equation}exp\left(i \phi\, 2 \pi\, q \right)= exp \left( \, \frac{2 \pi i q}{a}\, \,\cdots\, \frac{2 \pi i q}{a},\,  - \frac{2 \pi i
q}{N-a}\, \cdots \, - \frac{2 \pi i q}{N-a}\right),\end{equation}
an element of the center of $SU(a)\times SU(N-a)$, and come back
trough the $SU(a)\times SU(N-a)$ part (modulo an integer number L
of winding inside the U(1)). It follows that each value of the
magnetic charge uniquely corresponds to an element of
$\Pi_1(\tilde{H})$. The Dirac quantization condition is always
satisfied in terms of the minimal charge \cite{b}\cite{Preskill}.
This statement can be shown to hold for all the monopoles
corresponding to the symmetry breakings listed in the table of
section 4.3\footnote{\, We have checked this issue explicitly for
the non exceptional groups and for $G_2$. For $F_4$, $E_6$,
$E_7$ and $E_8$ is a conjecture.}. In section 4.2 we will study
the case of the $G_2$ group in detail.\newline In the cases where
$G$ is not simply connected (e.g. in the 't Hooft - Polyakov
solitonic solution
 $G = SO(3) \to U(1)$)  we must exclude the non contractible paths inside G and this fact restricts the allowed values for the magnetic charge.\\
The one-to-one correspondence between magnetic charges and
elements of $\pi_1(\tilde{H})$ allows to classify every
topological configuration in terms of the magnetic charge which is
defined in terms of the 't Hooft tensor (eq.(\ref{S})(\ref{S1})). 
The explicit construction of the tensor will be the main goal of the next section.

\section{The 't Hooft Tensor} \label{tensor}
\subsection{Construction}
The 't Hooft tensor is a gauge invariant tensor which coincides with
the residual abelian field strength in the unitary gauge. The magnetic field
associated to the i-th monopole is that of the residual gauge
group $U(1)^{i}$ generated by
$T_3^{\, i}$ . We can define the e.m. field  $A_{\mu}^{i}$ in terms of the gauge field $A_{\mu}^{\prime}$ in the
 unitary gauge as :
\begin{equation}
A_{\mu}^{i}=Tr(\phi^{i}_{0}\, A_{\mu}^{\prime})
\end{equation}  $\phi_{0}^i = \mu^{i}$, the fundamental weight (i = $1,\ldots,r$), identifies the monopole species.
If $b(x)$ is the gauge transformation bringing to a generic gauge
and $A_{\mu}$ the transformed gauge field \cite{Madore1}
\begin{equation}\left\{\begin{array}{cc} A_{\mu}^{\prime} = b A_{\mu} b^{-1} - \frac{i}{g}(\partial_{\mu}b) b^{-1}\\
{}\\ \phi^i_0 = b \phi^i b^{-1}
 \end{array}\right.\label{eq.14}\end{equation}
the e.m. field can be written as:
\begin{equation}
A_{\mu}^{i}=Tr( \phi^{i}( A_{\mu} + \Omega_{\mu}))
\end{equation}
where $\Omega_{\mu}= - \frac{i}{g}\, b^{-1}\partial_{\mu}b $. We
can rewrite the abelian field strength as
\begin{equation}F_{\mu\nu}^{i}=Tr(\phi^{i}\, G_{\mu \nu})+
i\, g\, Tr( \phi^{i} \, [ A_{\mu} + \Omega_{\mu}, A_{\nu}  +
\Omega_{\nu}])\label{eq55bis}\end{equation} Because of the
ciclycity of the trace only the part of $A_{\mu}-\Omega_{\mu}$
which does not belong to the invariance group of $\phi^i$
contributes. Indeed denoting for the sake of simplicity as
$V_{\mu}$ the vector $A_{\mu} + \Omega_{\mu}$
\begin{equation}Tr(\phi^i[ V_{\mu}, V_{\nu}]) = Tr\, ( V_{\nu}[ \phi^i,
V_{\mu}])= Tr\, ( V_{\mu}[ V_{\nu},
\phi^i])\label{a}\end{equation}To compute the second term in
eq.(\ref{eq55bis}) it proves convenient to introduce a projector
$P$ on the complement of the invariance algebra of $\phi^i$ . If
we write $V_{\mu}$ as
\begin{equation}V_{\mu} = \sum_{\vec{\alpha}} V_{\mu}^{\vec{\alpha}} E^{\vec{\alpha}} + \sum_j V_{\mu}^j
H^j\label{new}\end{equation} where the sum on $\vec{\alpha}$ is
extended to all positive and negative roots and the sum on $j$ on
all elements of Cartan algebra ($j=1,\ldots,r$), we can certainly
neglect the last term, which commutes with $\phi^i$. Moreover the
generic $E^{\vec{\alpha}}$ is part of the little group of $\phi^i$
whenever
\begin{equation}[\phi^i, E^{\vec{\alpha}}] = ( \vec{c}^{\, \, i} \cdot \vec{\alpha} )\, E^{\vec{\alpha}} = 0 \end{equation}
If instead $( \vec{c}^{\, \, i} \cdot \vec{\alpha} ) \neq 0$,
$E^{\vec{\alpha}}$ belongs to the complement. It is trivial to
verify that  projection on the complement $P^i\, V_{\mu}$ is given
by
\begin{equation}P^i V_{\mu} = 1 - \prod_{\vec{\alpha}}^{\prime} \left( 1 -  \frac{ [ \phi,[\phi,\, \, ]]
}{( \vec{c}^{\, \, i} \cdot \vec{\alpha} )^2}\right) V_{\mu}\label{P}
\end{equation}
where $[ \phi^i,\, \, \,   \,  ]\, V_{\mu} = [ \phi, V_{\mu} ]$
and the product $\prod_{\vec{\alpha}}^{\prime}$ run on the roots
$\vec{\alpha}$ such that $ \vec{c}^{\, \, i} \cdot \vec{\alpha}
\neq 0$ and only one representative is taken of the set of the
roots having the same value of $ \vec{c}^{\, \, i} \cdot
\vec{\alpha}$.\newline Indeed if any element $E^{\vec{\alpha}}$ in
eq.(\ref{new}) commutes with $\phi^i$, $ P^i E^{\vec{\alpha}} = (
1 - 1 ) E^{\vec{\alpha}} = 0$. If for any $E^{\vec{\alpha}}$
\begin{equation}[ \phi^i , E^{\vec{\alpha}}] = ( \vec{c}^{\, \, i} \cdot \vec{\alpha} ) E^{\vec{\alpha}} \, \, \, \, \, \, \, \, \, \, \, ( \vec{c}^{\, \, i} \cdot \vec{\alpha} ) \neq 0 \end{equation}
one of the factors $\left( 1 -  \frac{ [ \phi,[\phi \, \, ]] }{(
\vec{c}^{\, \, i} \cdot \vec{\alpha} )^2}\right)$ in the
definition eq.(\ref{P}) will give zero and $ P E^{\vec{\alpha}} =
E^{\vec{\alpha}}$.\newline In order to simplify the notation we
denote by $\lambda_I^i$ the different non zero values which $(
\vec{c}^{\, \, i} \cdot \vec{\alpha} )^2$ can assume and rewrite
$P^i V_{\mu}$ as
\begin{equation}P^i V_{\mu} = 1 - \prod_{I} \left( 1 -  \frac{ [ \phi,[\phi,\, \, ]]
}{ \lambda_I^i }\right) V_{\mu}\label{P2}
\end{equation}
Eq.(\ref{eq55bis}) can be rewritten as
\begin{equation}F_{\mu\nu}^{i}=Tr(\phi^{i}\, G_{\mu \nu})+
i g Tr( \phi^{i} \, [ P^i \left( A_{\mu} + \Omega_{\mu} \right),
A_{\nu}  + \Omega_{\nu}])\label{eq55}\end{equation} For our
purpose it is sufficient to project only one of the operators in
the commutator. By use of eq.(\ref{P2}) and recalling that
\begin{equation}D_{\mu} \phi^{\, i} = - i g [ A_{\mu} + \Omega_{\mu}\, ,\, \phi^{\, i}\, ]\end{equation}
the generalized 't Hooft tensor reads as
\begin{displaymath} F_{\mu \nu}^i = Tr ( \phi^i G_{\mu \nu} ) - \frac{i}{g}
\sum_{I} \frac{1}{\lambda_I^{i\, 2}  \,}\, \, Tr \left( \phi^i [D_{\mu} \phi^i, D_{\nu} \phi^i ] \right) +
\end{displaymath}
\begin{equation}
+\, \frac{i}{g} \sum_{I \neq J}\frac{1}{\lambda_{I}^{i\, 2}
\lambda_{J}^{i\, 2}}\, Tr \left( \phi^i [[D_{\mu} \phi^i, \phi^i],
[D_{\nu} \phi^i,\phi^i ]]\right) + ....\label{r1}\end{equation}
To
summarize we have to compute for each root $\vec{\alpha}$ the
(known) commutator $[ \phi^i , E^{\vec{\alpha}} ] = ( \vec{c}^{\,
\, i} \cdot \vec{\alpha} )\, E^{\vec{\alpha}}$ where
 $\phi^i$ are the fundamental weights associated to each simple root. This will give us the set of the values of $\lambda_I^i$ to insert into eq.(\ref{r1}).
For $SU(N)$ group $[\phi^i ,E_{\vec{\alpha}}]=(  \vec{c}^{\, \, i}
\cdot \vec{\alpha} ) E_{\vec{\alpha}}$ where $(  \vec{c}^{\, \, i}
\cdot \vec{\alpha}) = 0,\pm 1$, so the projector is simply
\begin{equation} P^i V_{\mu} = [\phi^i,[\phi^i, V_{\mu} ]]\end{equation}
and the 't Hooft tensor is the usual one
\begin{equation} F_{\mu \nu}^a = Tr ( \phi^a G_{\mu \nu} ) -
\frac{i}{g}Tr ( \phi^a [D_{\mu} \phi^a, D_{\nu} \phi^a ]
)\end{equation} For a generic group the projector is more
complicated and it can depend on the root chosen. Results are
listed in table of section 4.3.

\subsection{'t Hooft tensors for $G_2$ }

We now specialize the above results to the case of gauge group
$G_2$. It is natural to view $G_2 $ as a subgroup of $SO(7)$
\cite{berna1}. In fact $G_2$ is the subgroup of the $ 7 \times 7$
orthogonal matrices $\Omega$ which satisfy the relations
\begin{equation}T_{abc}= T_{def} \Omega_{da} \Omega_{eb} \Omega_{fc}\end{equation}
$T_{abc}$ is a totally antisymmetric tensor whose non-zero
elements are given by
\begin{displaymath}T_{127}= T_{154}=T_{235}= T_{264}=T_{374}=
T_{576} = 1\end{displaymath} According to section 2, we consider
the breaking of $G_2$ to a subgroup $SU(2) \times U(1)$. Dynkin
diagram of $G_2$ is depicted as follow
\begin{center}\begin{xy}
(0,0)*{};(60,-3)*\cir<7pt>{};(65,-3)*{}
**\dir3{-};(70,-3)*\cir<4pt>{}**\dir3{-}
\end{xy}\end{center}
\vspace{0.5cm} where the first circle corresponds to the simple
root $e_1$ = ( -2, 1, 1  ) and the second one to the simple root
$e_2$= ( 1, -1, 0 ). The residual invariance group is obtained by
erasing one of the two roots in turn. It's Dynkin diagram
consists of one single circle, which means $H = SU(2)$. The explicit
form of the generators of these residual $SU(2)$ subgroups is, in
the notation of \cite{berna1},
\begin{center}
$T_+^{(1)}$ = $\sqrt{2}$ (  $|1\rangle\langle2| - |5\rangle\langle 4|$ ) \, \, \, \, $T_{-}^{(1)}$ = $\sqrt{2}$ ( $|2\rangle\langle1| - |4\rangle\langle 5|$ )
\end{center}
\begin{center}
 $T_{3}^{(1)}$ = $ ( \,  |1\rangle\langle1| - |2\rangle\langle2| - |4\rangle\langle4| + |5\rangle\langle5|$ )
\end{center}
\begin{center}
$T_+^{(2)}$ = $\sqrt{2}|3\rangle\langle5| - \sqrt{2}|2\rangle\langle 6| - 2|7\rangle\langle 1| -  2|4\rangle\langle 7|$
\end{center}
\begin{center}
$T_-^{(2)}$ = $\sqrt{2}|5\rangle\langle3| - \sqrt{2}|6\rangle\langle 2| - 2|1\rangle\langle 7| -  2|7\rangle\langle 4|$
\end{center}
\begin{center}
 $T_{3}^{(2)}$ = $2 |1\rangle\langle1| - |2\rangle\langle2| - |3\rangle\langle3| - 2 |4\rangle\langle4| + |5\rangle\langle5| + |6\rangle\langle6|$
\end{center}
\begin{itemize}
\item If we break the simple root $e_1$ we have as little group $SU(2)\times U(1)$ and the corresponding maximal weight reads
\begin{equation}\phi_0^{(1)} = \, \texttt{diag} \frac{1}{2}( 0 , - 1 , 1, 0, 1, - 1, 0) \end{equation}
The coefficients ($\lambda_I^{(1)})$ are equal to $ 1 , 4$. By
using eq.(\ref{r1})  't Hooft tensor reads :
\begin{displaymath} F_{\mu \nu}^{(1)} = Tr ( \phi^{(1)} G_{\mu \nu} ) - \frac{5 i}{4 g}\,
 Tr \left( \phi^{(1)} [D_{\mu} \phi^{(1)}, D_{\nu} \phi^{(1)} ] \right) + \end{displaymath}\begin{equation}+ \frac{ i}{4 g}\, Tr \left( \phi^{(1)} [[D_{\mu} \phi^{(1)}, \phi^{(1)}], [D_{\nu}
\phi^{(1)},\phi^{(1)} ]]\right)\end{equation} More precisely the
invariance subgroup is $\frac{SU(2)\times U(1)}{Z_2}$. Indeed, if
we write $T_3^{(1)}$ as
\begin{equation}T_3^{(1)} = ( 1,-1,0,-1,1,0,0) = \phi_0^{(1)} + h\end{equation}
where $h$ is
\begin{equation}h = ( 1,-1/2,-1/2,-1,1/2,1/2,0) \end{equation}
we can see that when magnetic charge are even integers, the
corresponding loops wind only in the $U(1)$, while for odd integer
the loops travel partly  in $U(1)$, from identity to the
non-trivial element of the center of SU(2), and the rest in the
non-abelian $SU(2)$ subgroup. \item If we break the other simple
root $e_2$ we have as little group $SU(2)'\times U(1)$ and the
correspondent maximal weight reads
\begin{equation}\phi_0^{(2)} =  \texttt{diag} \frac{1}{6}( -1 , -1 , 2, 1, 1, -2, 0 ) \end{equation}
with $(\lambda_I^{(2)\,}) = 1, 4, 9$. These values of coefficients
give us a 't Hooft tensor of the form
\begin{displaymath} F_{\mu \nu}^2 = Tr (\phi^{(2)} G_{\mu \nu}) - \frac{49 i}{36 g}\,
 Tr \left(\phi^{(2)} [D_{\mu} \phi^{(2)}, D_{\nu} \phi^{(2)} ] \right) +\end{displaymath}\begin{displaymath}+
 \frac{7 i}{18 g}\, Tr \left( \phi^{(2)} [[D_{\mu} \phi^{(2)}, \phi^{(2)}], [D_{\nu}\phi^{(2)},\phi^{(2)} ]]\right)\end{displaymath}\begin{equation}- \frac{ i}{36 g}\, Tr \left( \phi^{(2)} [[[D_{\mu} \phi^{(2)}, \phi^{(2)}],\phi^{(2)}],[[D_{\nu}
\phi^{(2)},\phi^{(2)} ],\phi^{(2)}]]\right)
\end{equation}
\end{itemize}
Similarly to the previous case the residual gauge group is
$\frac{SU(2)\times U(1)}{Z_2}$ and for even charge loops wind only
on $U(1)$, while for odd charge loop run partly in $U(1)$ and the
rest in $SU(2)$.
\newpage
\subsection{'t Hooft tensor for generic group}
Below is a table of the symmetry breakings for all compact group.
In the second column we display the residual symmetry \footnote{We
indicate the breaking of the algebra but not the possible Z
factors of eq.(\ref{H}).\newline Notation $Spin(n)$ indicates the
universal covering group of $SO(n)$.}   $H \times U(1)$ which
uniquely determines  the corresponding values of $\lambda^I$
(third column) of eq.(\ref{P2}) and the associated 't Hooft
tensor. In the fourth column homotopies $\Pi_2(G/\tilde{H})$ with
$\tilde{H}$ defined in eq.(\ref{H}) are given.
\begin{small}
\begin{center}
\begin{tabular}{|c|c|c|c|}
\hline
$G$ & $ H \times U(1)\, \, \,$ & $\lambda_I$ & $\Pi_2(G/\tilde{H})$\\
\hline
\textbf{$SU(n)$} & $SU(n-m) \times SU(m) \times U(1)$ &  1  & \textbf{Z} \\
\hline
\textbf{$SO(2n+1)$} & $SO(2n-1) \times U(1)$ \, \, &  1   & \textbf{Z}\\
\hline
\textbf{$SO(2n+1)$} & $SO(2m+1) \times SU(n-m) \times U(1)$ \, \, &  1,4   & \textbf{Z}\\
\hline
\textbf{$SO(2n+1)$} & $SU(n) \times U(1)$ \, &  1,4  & \textbf{Z}/$Z_2$\\
\hline
\textbf{$SO(2n)$} & $SO(2n-2) \times U(1)$ &  1 & \textbf{Z} \\
\hline
\textbf{$SO(2n)$} & $SO(2m) \times SU(n-m) \times U(1)$ &  1,4 & \textbf{Z} \\
\hline
\textbf{$SO(2n)$} & $SU(n-2) \times SU(2) \times SU(2) \times U(1)$ \, &  1,4 & \textbf{Z}/$Z_2$\\
\hline
\textbf{$SO(2n)$} & $SU(n) \times U(1)$ \, &  1 & \textbf{Z}/$Z_2$\\
\hline
\textbf{$Sp(2n)$} & $Sp(2m) \times SU(n-m) \times U(1)$ &  1,4  & \textbf{Z}\\
\hline
\textbf{$Sp(2n)$} & $SU(n-1) \times SU(2) \times U(1)$ &  1,4  & \textbf{Z}\\
\hline
\textbf{$Sp(2n)$} & $SU(n) \times U(1)$  &  1 & \textbf{Z} \\
\hline
\textbf{$G_2$} & $SU(2) \times U(1)$ \, &  1,4,9  & \textbf{Z}\\
\hline
\textbf{$G_2$} & $SU(2)' \times U(1)$ \, &  1,4 & \textbf{Z}\\
\hline
\textbf{$F_4$} & $Sp(6) \times U(1)$ \, &  1,4 & \textbf{Z}\\
\hline
\textbf{$F_4$} & $SU(3) \times SU(2) \times U(1)$ \, &  1,4,9 & \textbf{Z}\\
\hline
\textbf{$F_4$} & $SU(3)' \times SU(2)' \times U(1)$ \, &  1,4,9,16 & \textbf{Z}\\
\hline
\textbf{$F_4$} & $Spin(7) \times U(1)$ \, &  1,4 & \textbf{Z} \\
\hline
\textbf{$E_6$} & $Spin(10) \times U(1)$ \, &  1 & \textbf{Z} \\
\hline
\textbf{$E_6$} & $SU(5) \times SU(2) \times U(1)$ \, &  1,4 & \textbf{Z}\\
\hline
\textbf{$E_6$} & $SU(6) \times U(1)$ \, &  1,4 & \textbf{Z}\\
\hline
\textbf{$E_6$} & $SU(3) \times SU(3) \times SU(2) \times U(1)$ \, &  1,4,9 & \textbf{Z}\\
\hline
\textbf{$E_7$} & $Spin(12) \times U(1)$ \, &  1,4 & \textbf{Z} \\
\hline
\textbf{$E_7$} & $SU(7)  \times U(1)$ \, &  1,4 & \textbf{Z}\\
\hline
\textbf{$E_7$} & $SU(6) \times SU(2) \times U(1)$ \, &  1,4,9  & \textbf{Z}\\
\hline
\textbf{$E_7$} & $SU(4) \times SU(3) \times SU(2) \times U(1)$ \, &  1,4,9,16 & \textbf{Z} \\
\hline
\textbf{$E_7$} & $SU(5) \times SU(3) \times U(1)$ \, &  1,4,9 & \textbf{Z}\\
\hline
\textbf{$E_7$} & $Spin(10) \times SU(2) \times U(1)$ \, &  1,4 & \textbf{Z}\\
\hline
\textbf{$E_7$} & $E_6 \times U(1)$ \, &  1 & \textbf{Z}\\
\hline
\textbf{$E_8$} & $Spin(14) \times U(1)$ \, &  1,4 & \textbf{Z}\\
\hline
\textbf{$E_8$} & $SU(8)  \times U(1)$ \, &  1,4,9 & \textbf{Z}\\
\hline
\textbf{$E_8$} & $SU(7) \times SU(2) \times U(1)$ \, &  1,4,9,16 & \textbf{Z}\\
\hline
\textbf{$E_8$} & $SU(5) \times SU(3) \times SU(2) \times U(1)$ \, &  1,4,9,16,25,36 & \textbf{Z}\\
\hline
\textbf{$E_8$} & $SU(5) \times SU(4) \times U(1)$ \, &  1,4,9,16,25 & \textbf{Z} \\
\hline
\textbf{$E_8$} & $Spin(10) \times SU(3) \times U(1)$ \, &  1,4,9,16 & \textbf{Z} \\
\hline
\textbf{$E_8$} & $E_6 \times SU(2) \times U(1)$ \, &  1,4,9 & \textbf{Z}\\
\hline
\textbf{$E_8$} & $E_7 \times U(1)$ \, &  1,4 & \textbf{Z}\\
\hline
\end{tabular}
\end{center}
\end{small}

\section{Discussion}

The experimental limits on the observation of free quarks in
nature indicate that confinement is an absolute property, in the
sense that the number of free quarks is strictly zero due to some
symmetry. Deconfinement is a change of symmetry. Since color is an
exact symmetry, the only way to have an extra symmetry, which can
be broken, is to look for a dual description of QCD. The extra
degrees of freedom are infrared modes related to boundary
conditions. This is a special case of the so called geometric
Langlands program of ref.\cite{Gukov:2006jk}.\newline The relevant
homotopy in 3+1 dimensions is a mapping of the two dimensional
sphere $S_2$ at spatial infinity onto the group. The homotopy
group is thus $\Pi_2$, configuration are monopoles
\cite{'tHooft:1974qc}\cite{d1} and the quantum numbers magnetic
charges.\newline For a generic gauge group of rank $r$ there exist
$r$ different magnetic charges $Q^a$ labelling the dual states.
The existence of magnetic charges implies a violation of Bianchi
identities by the abelian gauge field coupled to them. The gauge
invariant abelian field strength coupled to $Q^a$ is known as 't
Hooft tensor. In this paper we analyzed monopoles in a generic
compact gauge group and we explicitly constructed the
corresponding 't Hooft tensor.

\vspace{7.5cm}
\textbf{Aknowledgements}
\vspace{0.2cm}

We are very grateful to K. Konishi, F. Lazzeri, R. Chiriv\'{\i},
G. Cossu, M. D'Elia, V. Ghimenti, L. Ferretti and W. Vinci for
useful discussions. We also thank the Galileo Galilei Institute of
INFN for the hospitality during the workshop
\emph{"Non-Perturbative Methods in Strongly Coupled Gauge
Theories"}, where most of this work was accomplished.

\end{document}